\documentclass[12pt,a4paper]{article}
\usepackage[utf8]{inputenc}
\usepackage{amsmath,amssymb,amsfonts}
\usepackage{graphicx}
\usepackage{epstopdf}
\usepackage{hyperref}
\usepackage{cite}
\usepackage{geometry}
\usepackage{braket}
\usepackage{physics}
\usepackage{bm}
\usepackage{authblk}
\usepackage{indentfirst}

\title{\textbf{Non-Markovian Entropy Dynamics in Living Systems from the Keldysh Formalism}}

\author[1]{Feiyi Liu\thanks{Corresponding author: \texttt{fyliu@cxtc.edu.cn}}}
\author[1]{Min Guo}
\author[2]{Hongwei Tan}
\author[3]{Yang Wang}

\affil[1]{School of Physics, Electrical and Energy Engineering, Chuxiong Normal University, Chuxiong, 675000, China}
\affil[2]{School of Science, Hunan Institute of Technology, Hengyang 421002, China}
\affil[3]{School of Information Science and Engineering, Shandong Institute of Petroleum and Chemical Technology, Dongying, 257061, China}
\date{\today}

\begin{document}

\maketitle

\begin{abstract}
Living systems are open nonequilibrium systems that continuously exchange energy, matter, and information with their environments, leading to stochastic dynamics with memory and active fluctuations. In this study, we develop a non-Markovian theoretical framework for the entropy dynamics of living systems based on the Keldysh functional formalism and stochastic thermodynamics. The approach naturally incorporates colored environmental noise, memory-dependent dissipation, and many-body interactions, yielding generalized Langevin dynamics and non-Markovian master equations. Within this framework we derive an exact frequency domain expression for the entropy production rate and show that violations of the fluctuation–dissipation relation provide a direct thermodynamic signature of active biological fluctuations. We further demonstrate that environmental memory enhances low-frequency fluctuations and entropy production, leading to critical slowing down near dynamical instability. These results provide a microscopic physical foundation for the entropy “bathtub’’ picture of living systems and connect entropy evolution with development, aging, and death in nonequilibrium dynamics.

\end{abstract}



\section{Introduction}\label{Introduction}

Living systems are fundamentally open, far-from-equilibrium entities characterized by continuous exchanges of energy, matter, and information with their surrounding environments \cite{Nicolis1977,Peliti2021,DeAngelis1987}. Understanding the thermodynamic principles governing their complex lifespans remains a central challenge in modern biophysics and stochastic thermodynamics \cite{Seifert2012,Fang2019}. The framework of stochastic thermodynamics provides a powerful approach to analyze such systems, as it systematically connects microscopic fluctuating dynamics with macroscopic thermodynamic observables, even far from equilibrium. This perspective has revealed how physical constraints, such as thermodynamic uncertainty relations, impose fundamental trade-offs between the precision, speed, and energetic cost of biological functions, ranging from molecular motors to cellular sensing~\cite{Barato2015}. However, a comprehensive thermodynamic theory of life must not only describe these constraints but also explain the characteristic trajectory of living systems from birth through maturity to death \cite{Hayflick2007,Gladyshev2015,Gnesotto2018}. 

Recent work has increasingly explored how nonequilibrium thermodynamic principles shape the life trajectories of biological systems \cite{Seifert2012,Denisov2024,Brodeur2025}. Living organisms operate as open dissipative systems that continuously exchange energy and entropy with their environments, and stochastic thermodynamic frameworks have begun to clarify how energy dissipation constrains biological development and aging. Under far-from-equilibrium conditions, mechanisms such as maximum entropy production have been proposed to explain the emergence of self-replicating chemical systems and the growth of biological complexity during evolution \cite{Sawada2025}. At larger biological scales, dissipative scaling theories suggest that organismal growth and aging are regulated by energetic constraints on entropy production across biological hierarchies \cite{Biosystems2024}. At smaller scales, thermodynamic analyses relate epigenetic aging to changes in chromatin organization \cite{Singh2025}, while studies of anomalous diffusion and critical aging dynamics reveal universal nonequilibrium signatures such as weak ergodicity breaking and long-time correlations in complex systems \cite{Hu2025,Henkel2025}. These studies highlight the central role of nonequilibrium thermodynamics in governing biological development, maturation, and aging, motivating theoretical frameworks that describe the evolution of thermodynamic quantities throughout the lifespan of living systems.

Within this context, the entropy bathtub model recently formulated within a Markovian framework describes the lifespan of an organism as a thermodynamic trajectory consisting of three stages~\cite{Fornalski2026}. Entropy decreases during development, remains approximately constant in a nonequilibrium steady state during maturity, and increases irreversibly during aging. This model provides a unified thermodynamic description of the biological life cycle within stochastic thermodynamics. However, the Markovian formulation assumes memoryless dynamics and therefore cannot capture the strong memory effects intrinsic to biological systems, including epigenetic inheritance, delayed regulatory responses, and cumulative damage processes. These limitations motivate the extension of the entropy bathtub framework to a non-Markovian nonequilibrium description.

A natural theoretical framework for treating nonequilibrium open systems with memory is the Keldysh nonequilibrium field theory \cite{Schwinger1951,Keldysh1965,Kamenev2011,Sieberer2016,Thompson2023}. By formulating dynamics on a closed time contour, this approach provides a unified path-integral description of the system and its environment, allowing environmental degrees of freedom to be systematically integrated out in the spirit of the influence functional formalism \cite{Feynman1963,Caldeira1983}. The resulting effective action naturally contains nonlocal temporal kernels that encode dissipation, fluctuations, and memory effects beyond the Markov approximation \cite{Choudhury2024,Wu2024}. Such Keldysh approaches have recently emerged as powerful tools for analyzing strongly driven and non-Markovian open quantum systems and their critical nonequilibrium dynamics. These properties make the Keldysh framework particularly suitable for constructing a non-Markovian formulation of the entropy bathtub model.

In this work, we reformulate and extend the entropy bathtub concept within the Keldysh non-Markovian framework, which treats the system and environment on equal footing and naturally captures memory effects and correlations arising from microscopic couplings. Inspired by our previous studies of Keldysh field theory in quantum dissipation systems \cite{Liu2026}, we investigate the impact of colored noise, non-Gaussian fluctuations, quantum coherence, and many-body interactions on the standard entropy bathtub profile. Our main contribution is not merely to extend known non-Markovian results to a biological context, but to provide a unified Keldysh-based open-system framework that connects microscopic system--environment coupling, memory kernels, generalized Langevin dynamics, non-Markovian master equations, entropy production, and fluctuation--dissipation violation within a single formalism. This framework also allows us to derive exact expressions for non-Markovian entropy production, analyze critical relaxation dynamics, and examine the evolution of system--environment entanglement. At the same time, the biological scenarios discussed below are intended as coarse-grained and illustrative interpretations of the formalism, rather than as quantitative descriptions of specific organisms or datasets.

This paper is organized as follows. Section~\ref{Formalism} presents the theoretical framework, including the Keldysh formalism, the derivation of non-Markovian memory kernels, the non-Markovian entropy bathtub model in Section~\ref{Bathtub}, and generalized fluctuation–dissipation relations in Section~\ref{Fluctuation}. The physical implications are discussed in Section~\ref{Noise} on colored noise and non-Gaussian fluctuations, Section~\ref{Interaction} on many-body interactions, and Section~\ref{Exact} on entropy production and fluctuation theorems. Section~\ref{Aging} analyzes the critical dynamics of aging and death, Section~\ref{Response} derives non-Markovian response functions, and Section~\ref{Evolution} examines the evolution of system–environment entanglement entropy. Section~\ref{Conclusion} summarizes our findings.

\section{Theory}
\subsection{Keldysh Formalism for Open Living Systems}\label{Formalism}
To describe a living system within the framework of an open quantum system in the Keldysh formalism, the world can be partitioned into the system of interest and its surrounding environment. The system may represent a biomolecule, a cell, or a regulatory network, while the environment includes the surrounding cytoplasm, extracellular matrix, or a thermal bath. In this study, we adopt natural units with $\hbar = k_B = 1$ to better explore the statistical properties of the system. The total Hamiltonian can be written as \cite{Breuer2002,Weiss2012}
\begin{equation}
H_{\text{tot}} = H_S + H_B + H_I,
\label{Hamiltonian}
\end{equation}
where each term carries a distinct physical and biological interpretation.

The system Hamiltonian is given as
\begin{equation}
H_S = \sum_x E_x \ket{x}\bra{x},
\end{equation}
which describes the internal degrees of freedom of the biological entity under consideration. The discrete states $\ket{x}$ represent distinct conformational or functional states of the system, such as folded and unfolded configurations of a protein, different gene expression levels, or distinct phenotypic states of a cell. The energies $E_x$ characterize the stability of each state and may depend on biochemical and physical factors, including chemical gradients, membrane potentials, or mechanical stresses. Within this coarse-grained description, transitions between these states correspond to biological processes such as protein folding, ligand binding, or cellular differentiation.

The bath Hamiltonian of Eq.~\eqref{Hamiltonian} can be formulated as \cite{Caldeira1983,Leggett1987}
\begin{equation}
H_B = \sum_k \omega_k b_k^\dagger b_k,
\label{H_B}
\end{equation}
which models the surrounding environment as a collection of harmonic oscillators. Here, $b_k^\dagger$ and $b_k$ are the creation and annihilation operators of the $k$th harmonic bath mode, respectively, and $\omega_k$ denotes the corresponding mode frequency. These bath modes provide an effective coarse-grained description of collective environmental fluctuations in biological media. For example, for a protein embedded in a fluctuating aqueous environment, a bath mode may represent a collective fluctuation of the local hydration shell or density field surrounding the protein. In this case, the operators can be written as \cite{Breuer2002,Weiss2012}
\begin{equation}
b_k=\frac{1}{\sqrt{2}}(Q_k+iP_k), \qquad
b_k^\dagger=\frac{1}{\sqrt{2}}(Q_k-iP_k),
\end{equation}
where $Q_k$ denotes the amplitude of the corresponding environmental fluctuation and $P_k$ is its conjugate momentum, satisfying $[Q_k,P_k]=i$. This makes explicit that each term in Eq.~\eqref{H_B} represents an effective environmental fluctuation mode in a biologically relevant setting. Such coarse-grained environmental modes are, at least in principle, connected to experimentally accessible fluctuation spectra, for example through scattering-based measurements, microrheology, or electrical fluctuation measurements, depending on the specific biological medium \cite{Abbasi(2023),Nolte(2024),White(2000)}. The cellular environment, consisting of water molecules, ions, lipids, and other molecular constituents, possesses a large number of degrees of freedom that undergo small fluctuations around equilibrium. The harmonic oscillator bath captures two essential environmental effects: dissipation through energy exchange with the system and stochastic noise arising from thermal fluctuations. These features are precisely those required to describe thermal fluctuations in living cells.

The interaction Hamiltonian describes how the biological subsystem exchanges energy and information with its surrounding environment,
\begin{equation}
H_I = \sum_x \ket{x}\bra{x} \otimes B_x, \quad \text{with} \quad B_x = \sum_k g_{x,k}(b_k + b_k^\dagger).
\label{interaction}
\end{equation}
The coupling constants $g_{x,k}$ depend on the state of the system, reflecting that different conformations or functional configurations interact differently with the surrounding medium. For instance, a protein in its unfolded state may expose hydrophobic residues, thereby modifying its interaction with surrounding water molecules compared with the folded configuration. Similarly, an open ion channel couples differently to the local electric field than a closed channel, and a dividing cell exerts mechanical forces on the extracellular matrix that differ from those generated by a quiescent cell. The operators $b_k + b_k^\dagger$ represent collective environmental displacements, such as local density or electric field fluctuations, which perturb the system.

The environmental influence is characterized by the bath correlation functions of the coupling operators. 
For the present interaction Hamiltonian the relevant correlations appear through the operator differences 
$B_x-B_{x'}$ associated with different system states:
\begin{equation}
C_{xx'}(\tau) = \langle (B_x(\tau)-B_{x'}(\tau))(B_x(0)-B_{x'}(0)) \rangle_B,
\end{equation}
where $B_x(\tau) = e^{iH_B \tau} B_x e^{-iH_B \tau}$ is the Heisenberg-picture operator with respect to the bath Hamiltonian, and the expectation value $\langle \cdot \rangle_B$ is taken with respect to the equilibrium bath density matrix $\rho_B$. These correlation functions quantify how environmental fluctuations at different times remain correlated. Rapidly decaying correlations correspond to effectively Markovian environments, whereas slowly decaying correlations generate memory effects in the system dynamics.

For the harmonic oscillator bath considered here, the correlation functions can be evaluated exactly and take the form \cite{Feynman1963,Caldeira1983,Weiss2012}
\begin{equation}
C_{xx'}(\tau) = \int_0^\infty \frac{d\omega}{\pi} J_{xx'}(\omega) \left[ \coth\left(\frac{\beta\omega}{2}\right)\cos(\omega\tau) - i\sin(\omega\tau) \right],
\label{Corr}
\end{equation}
where the spectral density is defined as \cite{Breuer2002,Weiss2012}
\begin{equation}
J_{xx'}(\omega) = \pi \sum_k (g_{x,k}-g_{x',k})^2 \delta(\omega-\omega_k),
\end{equation}
and $\beta = 1/T$ as inverse temperature. By appropriately specifying the spectral density, this framework can also capture more complex environmental properties such as colored noise and memory effects.

Integrating out the bath degrees of freedom using the Keldysh path integral formalism leads to an influence functional that encodes the full effect of the environment on the system \cite{Schwinger1961,Keldysh1965,Kamenev2011}. Because the bath enters the action quadratically, the integration can be performed exactly, yielding \cite{Feynman1963}
\begin{equation}
\mathcal{F}[\phi^+,\phi^-] = \exp\left\{ -\int_0^t ds \int_0^s du \sum_{x,x'} \phi_x^+(s) \mathcal{K}_{xx'}(s-u) \phi_{x'}^+(u) + \text{cross terms} \right\},
\end{equation}
where $\phi_x^\pm$ are fields on the forward and backward branches of the Keldysh contour. The memory kernel $\mathcal{K}_{xx'}(\tau)$ appearing in the influence functional 
is determined by the correlation function of the environmental degrees of freedom introduced above. 
For a harmonic oscillator bath with linear system--environment coupling, integrating out the bath variables yields a kernel that is proportional to the environmental correlation function \cite{Caldeira1983,Weiss2012},
\begin{equation}
\mathcal{K}_{xx'}(\tau) = g_x g_{x'} C_{xx'}(\tau),
\label{Gene_kernel}
\end{equation}
where $g_x$ denotes the system–bath coupling amplitude. This relation reflects the Gaussian nature of the harmonic oscillator bath. 
Because the environmental fluctuations are fully characterized by their two-point correlation functions, 
the influence functional generates an effective action in which the temporal correlations of the environment appear explicitly through the memory kernel. Consequently, the non-Markovian dynamics of the system are governed by the temporal structure of the environmental correlation function $C_{xx'}(\tau)$.

The influence kernel $\mathcal K_{xx'}(\tau)$ inherits the temporal structure of the environmental correlation function $C_{xx'}(\tau)$ and determines the effect of past environmental fluctuations on the present system dynamics.
The $\sin(\omega\tau)$ term, associated with the imaginary part, describes energy exchange with the environment and therefore corresponds to dissipative dynamics. The $\coth(\beta\omega/2)$ factor in the real part accounts for both thermal and quantum fluctuations, which act as stochastic noise on the system. The explicit dependence on the time difference $\tau$ implies that the system dynamics depend on their past history, a characteristic feature of non-Markovian dynamics. Such memory effects are particularly relevant in biological systems, where past events, such as epigenetic modifications or accumulated damage, can influence future behavior.

In the classical high-temperature limit, the Keldysh kernel reduces to a real memory-friction kernel linked to generalized Langevin dynamics, with its decay time determined by environmental correlation functions that reflect biological memory. This microscopic origin of history-dependent dynamics lays the groundwork for the non-Markovian entropy bathtub examined in the following section.

\subsection{Non-Markovian Entropy Bathtub}\label{Bathtub}

Memory effects in biological environments can significantly modify the time evolution of entropy in living systems. Building on the memory kernel derived from the Keldysh formalism, we now examine the consequences of non-Markovian dynamics for the entropy bathtub. To this end, we consider a one-dimensional chain of discrete states representing successive developmental stages along an illustrative life-history trajectory, with transitions allowed only between neighboring states. The dynamics are governed by the generalized master equation (GME) \cite{Nakajima1958, Zwanzig1960, Breuer2002}:
\begin{equation}
\frac{\partial p_x(t)}{\partial t} = \sum_{x' = x \pm 1} \int_0^t ds \left[ K_{xx'}(t-s) p_{x'}(s) - K_{x'x}(t-s) p_x(s) \right],
\label{GME}
\end{equation}
where $p_x(t)$ is the probability to be in state $x$ at time $t$, and the memory kernel $K_{xx'}(\tau)$ encodes the history dependence of transitions from state $x'$ to $x$. The transition kernel $K_{xx'}(\tau)$ arises from coarse-graining the microscopic influence kernel $\mathcal K_{xx'}(\tau)$. To capture essential non-Markovian features while maintaining numerical tractability, we adopt an exponential kernel inspired by the Debye spectrum of an Ohmic bath \cite{Caldeira1983, Weiss2012}:
\begin{equation}
K_{xx'}(\tau) = \frac{k_{xx'}}{\tau_c} e^{-\tau/\tau_c}, \quad \text{for } |x-x'|=1.
\label{ex_ker}
\end{equation}
This form corresponds to the classical high-temperature limit $ T \gg \omega$ of the general Keldysh kernel of Eq.~\eqref{Corr} and \eqref{Gene_kernel}, where the imaginary (quantum) part vanishes and the kernel reduces to a real memory-friction kernel~\cite{Weiss2012}. 
In this limit the GME retains a probabilistic interpretation, and the purely real kernel isolates memory effects from quantum coherence. This simplification arises because the high-temperature regime $T\gg\omega$ suppresses phase coherence between system states, so that environmental fluctuations act effectively as classical noise. As a result, the dynamics can be consistently described in terms of transition probabilities between discrete states without introducing off-diagonal density-matrix elements. The memory time $\tau_c$ controls the extent of history dependence: $\tau_c \to 0$ recovers the Markovian limit $K_{xx'}(\tau) = k_{xx'}\delta(\tau)$, while finite $\tau_c$ introduces non-Markovian dynamics. In this limit the present formulation reduces to the Markovian entropy-balance ``bathtub'' model of living systems proposed in Ref.~\cite{Fornalski2026}, which therefore provides the baseline framework extended here by the inclusion of memory effects and temporally correlated fluctuations. The baseline rates $k_{xx'}$ satisfy detailed balance with the thermal bath at temperature $T$, i.e., $k_{x+1,x}/k_{x,x+1} = e^{-\beta(E_{x+1}-E_x)}$.

The system is initialized in thermal equilibrium, $p_x(0) \propto e^{-\beta E_x}$, with equilibrium rates $k_{xx'}^0$ satisfying detailed balance. At $t=0$, a time-dependent driving force is applied to bias forward transitions, mimicking the developmental process that reduces entropy during growth. Concretely, for nearest-neighbor transitions we write
\begin{equation}
k_{x+1,x}(t)=k_{x+1,x}^{0}e^{+\delta(t)}, \qquad
k_{x,x+1}(t)=k_{x,x+1}^{0}e^{-\delta(t)},
\label{driven_rates}
\end{equation}
so that the external driving enhances forward transitions and suppresses backward ones while preserving positivity of the rates. The corresponding non-Markovian transition kernel is therefore
\begin{equation}
K_{x\pm1,x}(\tau;t)=\frac{k_{x\pm1,x}(t)}{\tau_c}e^{-\tau/\tau_c}.
\label{driven_kernel}
\end{equation}
In this way, the driving acts directly on the amplitudes of the transition kernel, breaks detailed balance, and generates a probability current toward lower-entropy states during development.

For numerical simulations, we adopt a rectangular driving protocol as a minimal proof-of-principle choice, namely $\delta(t)=\delta_0$ during development and maturity and $\delta(t)=0$ thereafter. We emphasize, however, that this is an idealized approximation. In a more realistic biological setting, one expects a stronger drive during development and a weaker sustaining drive during maturity. This situation can be represented by a piecewise protocol,
\begin{equation}
\delta(t)=
\begin{cases}
\delta_{\mathrm{dev}}, & 0<t<t_{\mathrm{dev}},\\
\delta_{\mathrm{mat}}, & t_{\mathrm{dev}}<t<t_{\mathrm{age}},\\
0, & t>t_{\mathrm{age}},
\end{cases}
\qquad \text{with} \qquad
\delta_{\mathrm{dev}}>\delta_{\mathrm{mat}}>0.
\label{piecewise_drive}
\end{equation}
Under such a protocol, the entropy decrease during development is expected to be steeper, while the mature phase becomes flatter and less dissipative. If the mature-stage drive is too weak, the system drifts earlier toward the aging branch. By contrast, if it is too strong, the low-entropy mature state can be maintained longer, but at the cost of increased dissipation and entropy production. In the numerical examples below, unless otherwise stated, we use the simpler rectangular protocol in order to isolate the effect of memory from that of a more complicated time-dependent drive.

\begin{figure}[tbp]
\centering
\includegraphics[width=1\textwidth]{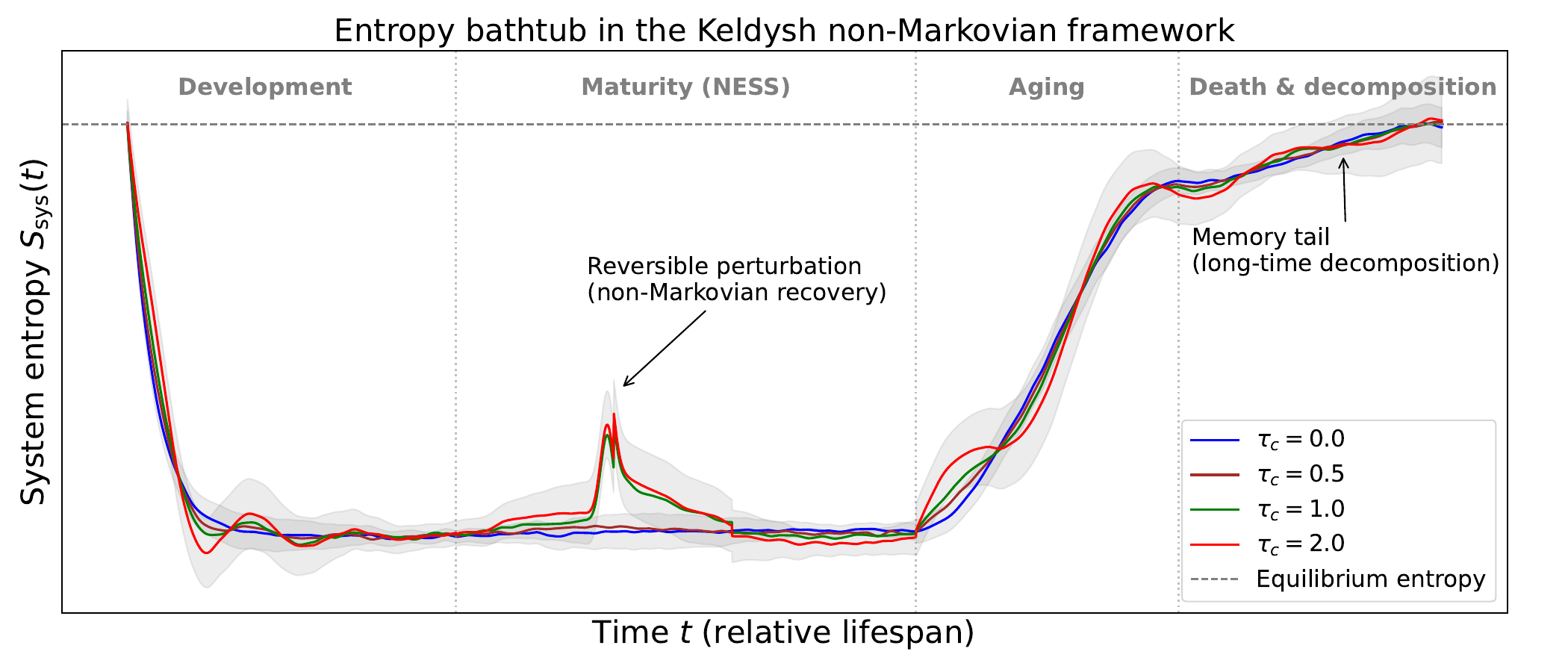}
\caption{Time evolution of system entropy $S_{\mathrm{sys}}(t)$ for different memory times $\tau_c = 0$ (blue), $0.5$ (brown), $1$ (green), and $2$ (red). Gray bands indicate fluctuation ranges for $\tau_c=0$ and $\tau_c=2$. The vertical dotted lines mark the boundaries between life stages (development, maturity, aging, death). A reversible perturbation (disease event) is shown for $\tau_c=1$ and $2$, illustrating non-Markovian recovery. The horizontal dashed line represents the equilibrium entropy of the environment.}
\label{fig_bathtub}
\end{figure}

To model a reversible perturbation during the mature phase, such as a disease event, we introduce an additional transient reduction of the sustaining drive. Specifically, during maturity we replace the driving term by
\begin{equation}
\delta(t)=\delta_{\mathrm{mat}}-\eta_{\mathrm{dis}}(t),
\label{disease_drive}
\end{equation}
where $\delta_{\mathrm{mat}}$ is the sustaining drive in the mature state and $\eta_{\mathrm{dis}}(t)$ is a localized pulse representing the disease event. A simple choice is
\begin{equation}
\eta_{\mathrm{dis}}(t)=A_{\mathrm{dis}}\,\Theta(t-t_{\mathrm{dis}})\,\Theta(t_{\mathrm{dis}}+\Delta t-t),
\label{disease_pulse}
\end{equation}
with amplitude $A_{\mathrm{dis}}>0$, onset time $t_{\mathrm{dis}}$, and duration $\Delta t$. The disease event is therefore incorporated into the generalized master equation through the same time-dependent transition rates and memory kernel defined in Eqs.~\eqref{driven_rates} and \eqref{driven_kernel}. In this case, the disease pulse temporarily weakens the forward ordering transitions and enhances the backward disordering transitions. Figure~\ref{fig_bathtub} shows the resulting entropy trajectory for several values of the memory time $\tau_c$.

The GME can be solved numerically via direct integration with discretized time steps, where convolution integrals are evaluated by storing the full history of probability distributions. Contributions beyond a few $\tau_c$ are negligible because of the exponential decay of the kernel. Ensemble averages over multiple independent realizations yield smooth entropy curves and fluctuation estimates. Figure~\ref{fig_bathtub} presents the time evolution of the system entropy \cite{Shannon1948,Seifert2012}
\begin{equation}
S_{\text{sys}}(t) = - \sum_x p_x(t) \ln p_x(t),
\label{entrop_bath}
\end{equation}
for four memory times: $\tau_c = 0$ (Markov limit, blue), $\tau_c = 0.5$ (brown), $\tau_c = 1$ (green), and $\tau_c = 2$ (red). The curves reveal several features that distinguish non-Markovian dynamics from the Markovian baseline.

During the developmental phase, the entropy decreases as the system is driven toward a more ordered state. For $\tau_c = 0$, the decline is smooth and approximately exponential. With finite memory, however, the decrease is modulated by small-amplitude oscillations that reflect the delayed response to the driving force: the system's current state is influenced not only by the instantaneous drive but also by its own past configurations. The amplitude of these oscillations increases with $\tau_c$, as the memory kernel extends further into the past.

In the mature steady state, where the driving force is maintained, the entropy attains a plateau. In the Markovian case, this plateau also exhibits stochastic fluctuations due to probabilistic transitions between states, as expected for a memoryless nonequilibrium steady state. Non-Markovian dynamics, however, introduce temporally correlated fluctuations together with a slow drift, visible in the shaded bands representing the standard deviation over realizations as gray bands for $\tau_c=0$ and $\tau_c=2$. These fluctuations arise because, although the average probability distribution is stationary, the two-time correlations encoded in the memory kernel sustain ongoing dynamical activity with history dependence. The mean entropy level also shifts slightly upward with increasing $\tau_c$, indicating that memory effectively renormalizes the transition rates and imposes an additional entropic cost to maintain the steady state.

Upon cessation of the driving force, the system enters the late relaxation regime, and the entropy rises toward its equilibrium value indicated by the horizontal dashed line. For $\tau_c = 0$, the increase is initially linear and then saturates exponentially, and a knee-like crossover may already be present in the Markovian description. In contrast, finite memory modifies the detailed shape of this regime: the upward curvature becomes more pronounced and the entropy approaches equilibrium more slowly, leaving a long tail.In this sense, non-Markovian effects do not necessarily introduce a qualitatively new late-stage feature, but rather alter its temporal structure by enhancing history-dependent relaxation. The late-time tail increases with $\tau_c$, indicating that systems with longer memory retain traces of their previously organized state for extended periods.

To further illustrate the impact of memory, Figure.~\ref{fig_bathtub} also includes an example of a reversible perturbation during the mature steady state, modeled by the transient pulse $\eta_{\mathrm{dis}}(t)$ introduced in Eqs.~\eqref{disease_drive} and \eqref{disease_pulse}, for $\tau_c = 1$ and $\tau_c = 2$. The perturbation produces a sharp entropy increase followed by a slow, non-Markovian recovery with a long tail, demonstrating that memory effects prolong the influence of transient stressors.
These numerical results show that the entropy bathtub is not a universal curve but carries the fingerprint of the system's microscopic memory. The transition kernel $K(\tau)$ arising from coarse-graining the microscopic influence kernel $\mathcal K(\tau)$ directly shapes the dynamics across all life stages. Specifically, the Markovian curve is smooth and featureless, whereas non-Markovian curves exhibit oscillations, drift, and an accelerated aging rise. The inset illustrates the memory kernels for the three finite-memory cases, demonstrating that the width of $K(\tau)$ governs these effects.These memory-induced modifications to the entropy trajectory have profound thermodynamic consequences, which can be systematically characterized through the frequency-dependent fluctuation-dissipation ratio.

\subsection{Generalized Fluctuation-Dissipation Relation and Its Violation in NESS}\label{Fluctuation}

The fluctuation–dissipation theorem provides a fundamental relation between spontaneous fluctuations and linear response in systems at
thermal equilibrium. These memory effects fundamentally alter how systems respond to external perturbations and how spontaneous fluctuations drive their dynamics \cite{Marconi(2008), Baiesi(2013),Puglisi(2017)}. The fluctuation-dissipation theorem provides a quantitative link between these two aspects, response and fluctuations, in equilibrium. Its violation therefore serves as a direct measure of how far a living system is driven away from equilibrium, offering a frequency-resolved probe of its non-equilibrium steady state and the underlying memory mechanisms.

A related experimentally motivated application of non-Markovian fluctuation--dissipation violation was recently discussed by Abbasi et al. \cite{Abbasi(2023)}, who studied active viscoelastic biomatter and analyzed frequency-dependent deviations from the fluctuation--dissipation theorem using a classical non-Markovian model of coupled degrees of freedom. In the classical high-temperature limit, our present formulation similarly reduces to a real memory-kernel description, which makes the connection with that class of models transparent. The main difference is that our approach starts from a quantum-mechanical Keldysh formalism for an open system and derives, within a unified framework, the corresponding non-Markovian dynamics, entropy production, and generalized fluctuation--dissipation relation for evolving biological systems.

In equilibrium, the fluctuation-dissipation theorem connects the symmetric correlation function $S_{AB}(\omega)$ to the imaginary part of the response function $\chi_{AB}''(\omega)$ through \cite{Kamenev2011,Weiss2012}
\begin{equation}
S_{AB}(\omega) = \coth\left( \frac{\omega}{2 T} \right) \chi_{AB}''(\omega),
\label{eq:FDT_equilibrium}
\end{equation}
where $S_{AB}(\omega)$ denotes the Fourier transform of the symmetrized correlation function
\begin{equation}
S_{AB}(t)=\frac12\langle A(t)B(0)+B(0)A(t)\rangle,
\end{equation}
and $\chi_{AB}''(\omega)=\mathrm{Im}\,\chi^R_{AB}(\omega)$ is the imaginary part of the retarded response function. In the Keldysh formalism these quantities are naturally expressed in terms of Green's functions: the symmetric correlator is related to the Keldysh Green's function $G^K$, while the response function is determined by the retarded component $G^R$ \cite{Kamenev2011, Dyson1949}. Eq.~\eqref{eq:FDT_equilibrium} is therefore equivalent to the well-known equilibrium relation
\begin{equation}
G^K(\omega)=\coth\left(\frac{\omega}{2 T}\right)\left[G^R(\omega)-G^A(\omega)\right].
\end{equation}
This relation is directly connected to the environmental correlation functions $C_{xx'}(\tau)$ introduced in Eq.~\eqref{Corr}, since the Fourier transform of their symmetrized part determines the fluctuation spectrum entering $S_{AB}(\omega)$. Physically, Eq.~\eqref{eq:FDT_equilibrium} expresses a fundamental balance, according to which the spontaneous fluctuations of a system at temperature $T$ are precisely proportional to its energy dissipation when probed by an external force. In biological terms, this means that a protein's thermal fluctuations around its native state are directly related to its dissipation of energy when mechanically stretched, or that membrane voltage fluctuations are tied to the membrane's electrical impedance.

For a non-equilibrium steady state sustained by continuous energy input from metabolic activity or external driving, the delicate balance expressed by the fluctuation-dissipation theorem is fundamentally disrupted. To maintain their organized structure, such living systems dissipate energy and export entropy, rendering the equilibrium relation inapplicable. We quantify this departure using the generalized fluctuation-dissipation ratio as
\begin{equation}
X_{AB}(\omega) = \frac{S_{AB}(\omega)}{ \coth\left( \frac{\omega}{2 T_{\text{ref}}} \right) \chi_{AB}''(\omega)},
\label{eq:X_definition}
\end{equation}
where $T_{\text{ref}}$ is a reference temperature typically chosen as the ambient temperature of the environment. The deviation of $X_{AB}(\omega)$ from unity provides a frequency-resolved measure of the extent to which the system deviates from equilibrium. In this sense, the nonequilibrium contribution encoded in $X_{AB}(\omega)\neq 1$ may be interpreted as arising from active driving, entropy export, and memory effects, in line with generalized fluctuation-dissipation frameworks for evolving nonequilibrium systems \cite{Marconi(2008),Baiesi(2013),Puglisi(2017)}.
Using the Keldysh technique in Section~\ref{Formalism}, this ratio can be expressed directly in terms of the self-energies that encode system-environment interactions \cite{Kamenev2011}:
\begin{equation}
X_{AB}(\omega) =\frac{\Sigma^K(\omega)}{ \coth\!\left(\frac{\omega}{2 T_{\text{ref}}}\right)\Sigma''(\omega)},
\label{X_selfenergy}
\end{equation}
where $\Sigma^K(\omega)$ is the Keldysh component of the self-energy representing fluctuations and noise, and 
\begin{equation}
\Sigma''(\omega) = \mathrm{Im}\,\Sigma^R(\omega)
\end{equation}
is the dissipative part of the retarded self-energy. In equilibrium these components satisfy the fluctuation-dissipation relation
\begin{equation}
\Sigma^K(\omega)=\coth\!\left(\frac{\omega}{2 T}\right)\Sigma''(\omega),
\label{FDT_selfenergy}
\end{equation}
which guarantees $X_{AB}(\omega)=1$.

For the harmonic oscillator bath considered in Section~\ref{Formalism}, both $\Sigma^K(\omega)$ and $\Sigma^R(\omega)$ are determined by the same bath spectral density and are therefore closely related to the Fourier transform of the memory kernel $\mathcal{K}(\tau)$. In particular, the dissipative part $\Sigma''(\omega)$ is associated with the imaginary component of the retarded kernel $\tilde{\mathcal K}^R(\omega)$, while the Keldysh component $\Sigma^K(\omega)$ encodes the noise spectrum determined by the bath correlations. When the environment possesses memory or when the system is driven out of equilibrium, the balance between $\Sigma^K$ and $\Sigma''$ is modified, leading to deviations of $X_{AB}(\omega)$ from unity.

Eq~\eqref{X_selfenergy} can be recast into a more intuitive form by introducing an effective frequency-dependent temperature $T_{\text{eff}}(\omega)$ \cite{Cugliandolo2011}, defined through Eq.~\eqref{FDT_selfenergy} with $T=T_{\text{eff}}(\omega)$. Substituting this relation into Eq.~\eqref{X_selfenergy} yields
\begin{equation}
X_{AB}(\omega)=\frac{\coth\!\left(\frac{\omega}{2 T_{\text{eff}}(\omega)}\right)}{\coth\!\left(\frac{\omega}{2 T_{\text{ref}}}\right)}.
\label{X_Teff}
\end{equation}
Thus the fluctuation-dissipation ratio directly reflects the degree to which the effective temperature sensed by the system deviates from the reference temperature at each frequency. In equilibrium $T_{\text{eff}}(\omega)=T_{\text{ref}}$ and $X=1$, whereas out of equilibrium $T_{\text{eff}}(\omega)$ generally becomes frequency dependent, encoding the non-equilibrium character of the dynamics.

The frequency dependence of $X_{AB}(\omega)$ may be interpreted in a biologically suggestive but coarse-grained way. To illustrate this point, we compute $X_{AB}(\omega)$ for the exponential memory kernel defined in Eq.~\eqref{ex_ker}. The memory time $\tau_c$ introduces a characteristic frequency scale $\omega_c = 1/\tau_c$. To represent different life stages, we model the effective temperature $T_{\text{eff}}(\omega)$ in a manner consistent with the entropy dynamics shown in Fig.~\ref{fig_bathtub}. During development, the system actively builds order and suppresses fluctuations at intermediate frequencies. This behavior is represented by a dip in $T_{\text{eff}}(\omega)$ centered near $\omega_c$, which can be written as  \cite{Cugliandolo2011}
\begin{equation}
T_{\text{eff}}(\omega) =T_{\text{ref}}\left[1 - A e^{-(\omega-\omega_c)^2/2\sigma^2}\right],
\end{equation}
where $A=0.3$ and $\sigma=0.2\,\omega_c$. This form leads to $X_{AB}(\omega)<1$ in that frequency range.

In the mature steady state, $T_{\text{eff}}(\omega)$ remains close to $T_{\text{ref}}$ but exhibits small oscillations due to residual memory flow. Consequently, $X_{AB}(\omega)$ stays close to unity with weak undulations. During aging, low-frequency fluctuations become amplified as damage accumulates. This effect is modeled by introducing a peak at low frequencies \cite{Fodor2018},
\begin{equation}
T_{\text{eff}}(\omega) = T_{\text{ref}}\left[1 + \frac{B}{1+(\omega/\omega_c)^2}\right],
\end{equation}
where $B=0.5$. This form yields $X_{AB}(\omega)>1$ for $\omega \ll \omega_c$. For simplicity, all three stages are assumed to share the same memory time $\tau_c = 2$.
Therefore, the differences in $X_{AB}(\omega)$ arise primarily from the distinct effective temperature profiles, while the memory kernel sets the common dissipative background.

\begin{figure}[tbp]
\centering
\includegraphics[width=0.6\textwidth]{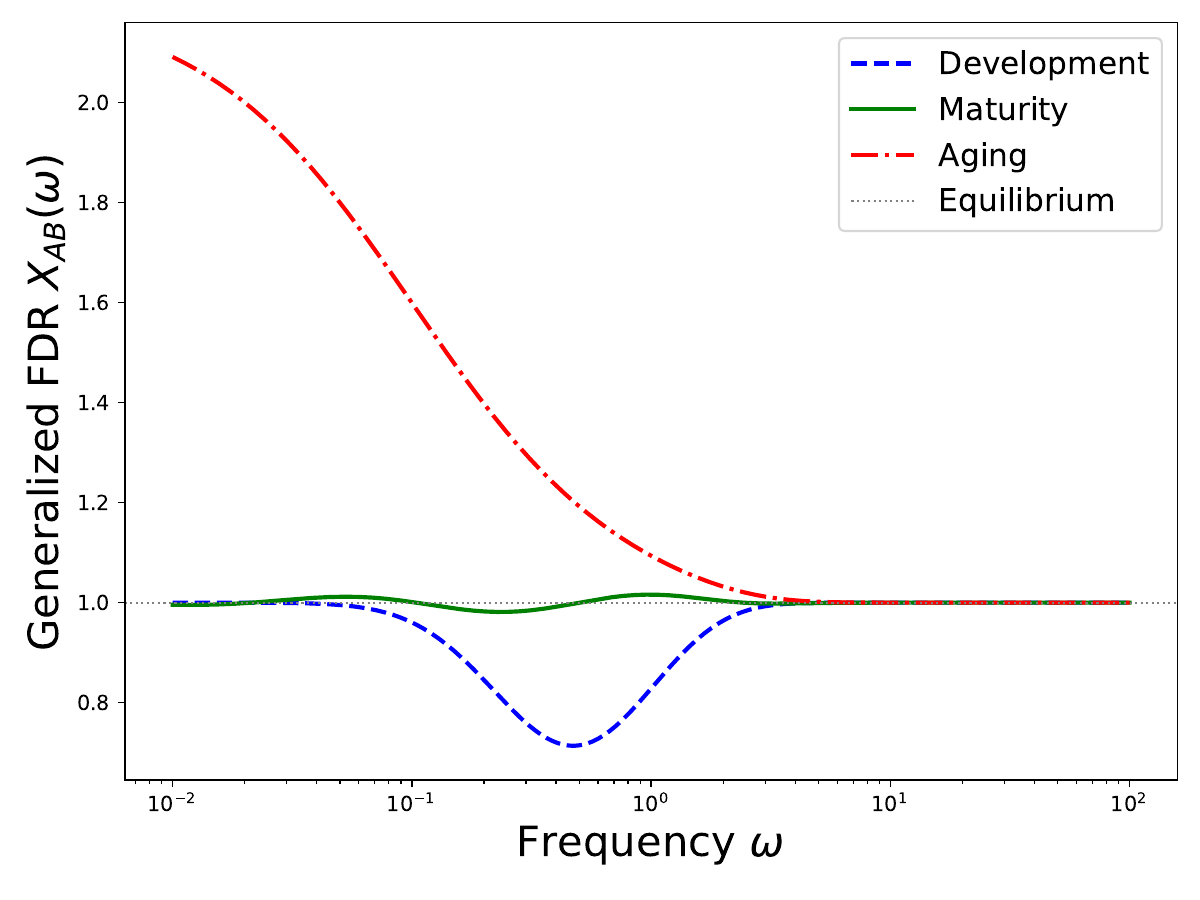}
\caption{Frequency-dependent fluctuation-dissipation ratio $X_{AB}(\omega)$ for the exponential memory kernel with $\tau_c=2$, computed from Eq.~\eqref{X_Teff} using effective temperature models representing different life stages. The blue dashed curve (development) shows a dip below unity around $\omega \sim 1/\tau_c$, indicating suppressed fluctuations. The green solid curve (maturity) remains close to $X_{AB}=1$ (the gray dotted line marks the equilibrium value) with small oscillations reflecting residual memory effects. The red dash-dotted curve (aging) exhibits a pronounced enhancement at low frequencies, signaling amplified noise and reduced dissipation efficiency. }
\label{fdratio}
\end{figure}

Figure~\ref{fdratio} displays the resulting $X_{AB}(\omega)$ curves computed from Eq.~\eqref{X_Teff} using these effective temperature models. In the developmental phase, the dip near the characteristic frequency $\omega_c = 1/\tau_c$ (here $\omega_c \approx 0.5$ since $\tau_c=2$) reflects the system's ability to channel energy into order-building processes while suppressing fluctuations at intermediate time scales. During maturity, $X_{AB}(\omega)$ remains close to unity across the entire frequency range, with very weak deviations. This behavior indicates that the system operates near a balanced non-equilibrium steady state where fluctuations and dissipation remain approximately matched. In aging, the pronounced enhancement of $X_{AB}(\omega)$ at low frequencies indicates that the system becomes increasingly sensitive to slow perturbations. Such behavior is consistent with reduced regulatory efficiency and the gradual accumulation of damage, which amplify long-time fluctuations and degrade the system's ability to dissipate them. These three regimes may therefore be viewed as distinct forms of fluctuation-dissipation violation: fluctuation suppression during development, near-balanced nonequilibrium dynamics in maturity, and fluctuation amplification during aging.

The generalized fluctuation-dissipation ratio therefore provides a direct experimental probe of nonequilibrium dynamics in living systems. By comparing spontaneous fluctuations with the response to weak perturbations, it offers a noninvasive method for quantifying thermodynamic irreversibility in biological processes. The frequency-dependent structure of $X_{AB}(\omega)$ reflects the temporal correlations encoded in the
memory kernel and therefore links microscopic non-Markovian dynamics to observable macroscopic response behavior.

\section{Discussion}

\subsection{Colored Noise and Non-Gaussian Fluctuations}\label{Noise}

Environmental fluctuations in living systems exhibit temporal correlations encoded in the bath correlation functions $C_{xx'}(\tau)$, which characterize the memory properties of the surrounding environment. After coarse-graining over microscopic bath degrees of freedom, these
correlations generate effective stochastic forces acting on collective variables of the biological system. The corresponding noise spectrum is
determined by the Fourier transform of the environmental correlations. When the environment possesses memory, the resulting fluctuations are
generally colored, meaning that their power spectrum depends on frequency. This contrasts with the ideal white-noise limit used in
Markovian descriptions, where correlations decay instantaneously. In biological systems colored noise is ubiquitous. For example, the viscoelastic cytoplasm often produces long-time correlations leading to $1/f^\alpha$ spectra at low frequencies. Such colored noise strongly influences molecular motor motion, protein folding kinetics, and chromosome dynamics.

A convenient description is provided by the generalized Langevin equation for a coarse-grained collective coordinate $q(t)$ \cite{Mori1965,Kubo1966,Zwanzig2001},
\begin{equation}
m\ddot{q}(t)=-V'(q)-\int_0^t \gamma(t-s)\dot q(s),ds+\xi(t),
\label{langevin}
\end{equation}
where the friction kernel $\gamma(\tau)$ arises from the coarse-grained influence of the environment. This kernel is determined by the microscopic environmental correlations encoded in the influence kernel $\mathcal K(\tau)$, which is derived within the Keldysh formalism. The stochastic force $\xi(t)$ represents environmental fluctuations, and its correlation function defines the noise spectrum \cite{Kubo1966,Callen1951}
\begin{equation}
\langle \xi(t)\xi(t') \rangle=S_\xi(t-t')=T\,\gamma(|t-t'|).
\label{noise_spectrum}
\end{equation}

In thermal equilibrium the fluctuation–dissipation theorem (FDT) requires \cite{Kubo1966}
\begin{equation}
S_\xi(\omega)=2T\,\mathrm{Re}\,\tilde{\gamma}(\omega),
\label{FDT}
\end{equation}
where $\tilde{\gamma}(\omega)$ is the Fourier transform of the kernel. In living systems driven by metabolic activity, this relation is generally violated, reflecting active nonequilibrium fluctuations. The linear response of the system to an external perturbation is characterized by the susceptibility
\begin{equation}
\chi(\omega)=\frac{1}{-m\omega^2-i\omega\tilde{\gamma}(\omega)+V''(q)}.
\label{chi_langevin}
\end{equation}

For a nonequilibrium steady state driven by colored noise,
the entropy production rate can be expressed in the frequency domain as \cite{Seifert2012, Harada(2005),Lippiello(2014),Golestanian(2025)}
\begin{equation}
\langle \dot S_{\mathrm{tot}} \rangle=\int_{-\infty}^{\infty}\frac{d\omega}{2\pi}\frac{\omega^2|\chi(\omega)|^2}{ T}\left[S_\xi(\omega)
-2  T\,\mathrm{Re}\,\tilde{\gamma}(\omega)\right],
\label{gene_entropy_production}
\end{equation}
which shows that entropy production arises from violations of the fluctuation–dissipation relation and will be discussed in detail in Section \ref{Exact}. This structure is closely related to the Harada-Sasa relation and its subsequent generalizations, which connect nonequilibrium dissipation to the violation of fluctuation-response balance \cite{Harada(2005)}. The enhanced low-frequency noise increases long-time fluctuations and therefore accelerates entropy production, providing a dynamical mechanism for the aging branch of the entropy bathtub. Beyond Gaussian fluctuations, biological systems often exhibit rare strong events such as transcriptional bursts, ion channel openings, or DNA damage. These processes generate non-Gaussian noise that cannot be characterized solely by the power spectrum.

In the Keldysh framework, the noise spectrum and friction kernel can be expressed in terms of the Keldysh and retarded
self-energies of the effective dynamics. Using $S_\xi(\omega)=\Sigma^K(\omega)$ and $\mathrm{Re}\,\tilde{\gamma}(\omega)=-\mathrm{Im}\,\Sigma^R(\omega)$, the entropy production rate can be written compactly as
\begin{equation}
\langle \dot S_{\mathrm{tot}} \rangle = \int_{-\infty}^{\infty}\frac{d\omega}{2\pi} \frac{\omega^2}{T} |G^R(\omega)|^2
\left[\Sigma^K(\omega)-2T\,\mathrm{Im}\,\Sigma^R(\omega)\right],
\label{Keldysh_entropy_production}
\end{equation}
where $G^R(\omega)=\chi(\omega)$ is the retarded Green function. This expression shows that entropy production is directly controlled by the violation of the fluctuation–dissipation relation at the level of the Keldysh self-energies, and may be viewed as the non-Markovian Keldysh counterpart of the Harada-Sasa class of relations.

Within the Keldysh influence functional formalism, such effects appear as higher-order cumulants of the environmental noise. The influence
functional can then be written as a cumulant expansion \cite{Kamenev2011}
\begin{equation}
\mathcal{F}
=\exp\Bigg[-\frac12\int dt_1 dt_2\, j(t_1) S_\xi(t_1-t_2) j(t_2)+\frac{i}{3!}\int dt_1 dt_2 dt_3C_3(t_1,t_2,t_3)j(t_1) j(t_2) j(t_3)+\cdots\Bigg],
\end{equation}
where $C_3$ denotes the third-order noise cumulant describing the skewness of the fluctuation statistics. These higher-order correlations modify the entropy production rate. To leading order, the correction takes the form
\begin{equation}
\langle \dot S_{\mathrm{tot}} \rangle=\langle \dot S_{\mathrm{tot}} \rangle_{\mathrm{Gauss}}+\frac{1}{6}\int dt_1 dt_2 dt_3\,
\Psi(t_1,t_2,t_3)C_3(t_1,t_2,t_3)+\mathcal O(C_4),
\end{equation}
where $\Psi(t_1,t_2,t_3)$ is a third-order response kernel determined by the system dynamics. Non-Gaussian fluctuations therefore introduce
intermittent entropy bursts that can manifest as abrupt changes in the entropy trajectory, corresponding to rare biological events such as sudden cellular damage or pathological transitions.

\subsection{Many-Body Interactions and Collective Behavior}\label{Interaction}

The collective dynamics of interacting units in many biological processes such as neuronal synchronization, collective cell migration, quorum sensing in bacterial colonies, and tissue-level morphogenesis cannot be understood from isolated single-particle behavior. To extend the Keldysh non-Markovian framework to many-body systems, consider a collection of $N$ interacting units with states $\{x_i\}=(x_1,x_2,\dots,x_N)$. The joint probability distribution$P(\{x_i\},t)$ obeys a generalized many-body master equation \cite{Gardiner2004}
\begin{equation}
\begin{aligned}
\frac{\partial P(\{x_i\},t)}{\partial t}=&\sum_i \int_0^t ds\sum_{x_i'}\Big[\mathcal{K}^{(i)}_{x_i x_i'}(\{x_{j\neq i}\},t-s)P(\{x_i'\},s)  \\
&-\mathcal{K}^{(i)}_{x_i' x_i}(\{x_{j\neq i}\},t-s)P(\{x_i\},s)\Big] \\
&+\sum_{i<j} \int_0^t ds\sum_{x_i',x_j'}\Big[\mathcal{K}^{(ij)}_{x_i x_j ,\, x_i' x_j'}(t-s)P(\{x_i',x_j',\dots\},s)  \\
&-\mathcal{K}^{(ij)}_{x_i' x_j' ,\, x_i x_j}(t-s)P(\{x_i,x_j,\dots\},s)\Big].
\end{aligned}
\end{equation}
The single-site memory kernel $\mathcal{K}^{(i)}$ captures the dependence of the transition dynamics of unit $i$ on both its own past history and the current configuration of neighboring units. The two-body kernel $\mathcal{K}^{(ij)}$ captures delayed interactions between units $i$ and
$j$, such as synaptic transmission delays in neural networks or transport-mediated signaling between cells.

For large systems it is often convenient to introduce coarse-grained collective fields. Let $\phi(\mathbf r,t)$ denote the local density of
cells, proteins, or signaling molecules. The coarse-grained dynamics can then be described by a non-Markovian stochastic field equation \cite{Tauber2014}
\begin{equation}
\partial_t \phi(\mathbf r,t)=D\nabla^2 \phi+\nabla\!\cdot\!\left[\phi(\mathbf r,t)\nabla\int U(\mathbf r-\mathbf r')\phi(\mathbf r',t)d\mathbf r'\right]
-\int_0^t ds\,\gamma(t-s)\partial_s\phi(\mathbf r,s)+\xi(\mathbf r,t).
\end{equation}
Here $U(\mathbf r-\mathbf r')$ represents an effective interaction potential arising from processes such as chemotaxis, adhesion, or electrical coupling. The stochastic force $\xi(\mathbf r,t)$ represents environmental and intrinsic fluctuations, whose correlations follow Eq.~\eqref{noise_spectrum}.

Within the Keldysh field-theoretical framework, the collective dynamics of the coarse-grained field can be characterized by the retarded Green function $G^R(\omega)$, which determines the linear response of the system to external perturbations. The response function satisfies the Dyson equation \cite{Dyson1949}
\begin{equation}
G^R(\omega)=\frac{1}{G_0^R(\omega)^{-1}-\Sigma^R(\omega)},
\label{Dyson_equation}
\end{equation}
where $G_0^R(\omega)$ represents the single-component dynamics and $\Sigma^R(\omega)$ is the retarded self-energy generated by interactions among different units. Collective coupling therefore renormalizes the effective relaxation dynamics through $\Sigma^R(\omega)$, which may soften near a collective transition and lead to long correlation times.

Upon introducing a global order parameter $\Phi(t)=\langle\phi(\mathbf r,t)\rangle$, the dynamics near collective transitions simplify to a memory-modified Ginzburg–Landau equation
\begin{equation}
\frac{d\Phi}{dt}=a(\mu)\Phi-b\Phi^3-\int_0^t K(t-s)\Phi(s)ds+\zeta(t),
\label{GL_memory}
\end{equation}
where $\mu$ is a control parameter such as coupling strength or cell density. The coefficient $a(\mu)$ changes sign at the critical point
$\mu_c$, signaling the onset of collective order.

Close to criticality the correlation time diverges as
\begin{equation}
\tau_c \sim |\mu-\mu_c|^{-\nu},
\label{corr_time_diverge}
\end{equation}
which strongly enhances fluctuations and dissipation. As a consequence, the entropy production rate exhibits critical scaling
\begin{equation}
\langle \dot S_{\mathrm{tot}} \rangle \sim |\mu-\mu_c|^{-\alpha},
\label{entropy_production}
\end{equation}
where the exponent $\alpha$ depends on the long-time tail of the memory kernel. Such dissipation bursts accompany collective biological
transitions, including neuronal synchronization, developmental pattern formation, and other large-scale reorganizations. Thus the many-body extension of the Keldysh non-Markovian framework connects microscopic memory kernels to macroscopic collective phenomena, providing a bridge between molecular fluctuations and the large-scale dynamics that shape the entropy bathtub of living systems.

\subsection{Exact Expression for Non-Markovian Entropy Production and Fluctuation Theorems}\label{Exact}

A central quantity in nonequilibrium thermodynamics is the entropy production associated with stochastic trajectories of the system. For systems driven by colored noise and memory-dependent dissipation, the trajectory probability must account for temporal correlations in the stochastic force. Consider a stochastic trajectory $\Gamma=\{q(t)\}_{0\le t\le \tau}$ generated by the generalized Langevin equation of Eq.~\eqref{langevin} \cite{Seifert2005}. The stochastic force $\xi(t)$ obeys Gaussian statistics with temporal correlations described by the colored-noise correlation function $S_\xi(\tau)$.The probability weight of a noise realization can therefore be written as \cite{Onsager1953}
\begin{equation}
\mathcal P[\xi]\propto\exp
\left[-\frac{1}{4}\int_0^\tau dt\int_0^\tau dt'\,\xi(t) S_\xi^{-1}(t-t') \xi(t')\right],
\end{equation}
where $S_\xi^{-1}$ is the inverse kernel satisfying
\begin{equation}
\int ds\,S_\xi(t-s)S_\xi^{-1}(s-t')=\delta(t-t').
\end{equation}
Through the Langevin equation, each noise realization uniquely determines a system trajectory $\Gamma$. The corresponding trajectory probability $\mathcal P[\Gamma]$ therefore inherits the same nonlocal temporal structure. Within stochastic thermodynamics, the total entropy production along a trajectory is defined as the logarithmic ratio between the probability of the forward trajectory $\Gamma$ and that of the time-reversed trajectory $\tilde{\Gamma}$,
\begin{equation}
\Delta S_{\mathrm{tot}}[\Gamma]=\ln\frac{\mathcal P[\Gamma]}{\mathcal P[\tilde{\Gamma}]} .
\end{equation}
For systems with non-Markovian dissipation in Kelysh theory, the entropy production rate can be expressed directly in terms of the self-energy components that characterize dissipation and fluctuations of the environment as given in Eq.~\eqref{Keldysh_entropy_production}. It shows explicitly that entropy production arises from violations of the fluctuation–dissipation theorem. The quantity $\Sigma^K(\omega)-2T\,\mathrm{Im}\,\Sigma^R(\omega)$ measures the violation of the fluctuation–dissipation relation and therefore determines the irreversible entropy production of the system. 

The trajectory-level definition of entropy production further implies universal fluctuation theorems, and in particular, the detailed fluctuation theorem takes the form
\begin{equation}
\frac{P(\Delta S_{\mathrm{tot}})}{P(-\Delta S_{\mathrm{tot}})}=e^{\Delta S_{\mathrm{tot}}},
\end{equation}
while integrating over all trajectories yields the integral fluctuation theorem
\begin{equation}
\langle e^{-\Delta S_{\mathrm{tot}}} \rangle = 1.
\end{equation}
These relations remain valid even in the presence of colored noise and memory effects because they rely only on the microscopic reversibility
of the stochastic dynamics.

In living systems, however, fluctuations are often strongly non-Gaussian due to intermittent biological events such as transcriptional bursts, ion-channel openings, metabolic activation, or sudden cellular damage. These processes produce large, sporadic entropy increments that dominate the tails of the entropy distribution.The statistics of such rare events can be described within a large-deviation framework. For long observation times $\tau$, the entropy production distribution takes the asymptotic form \cite{Evans1993}
\begin{equation}
P(\Delta S_{\mathrm{tot}})\sim\exp\left[-\tau\, I\!\left(\frac{\Delta S_{\mathrm{tot}}}{\tau}\right)\right],
\end{equation}
where $I(s)$ is the large-deviation rate function. Entropy bursts correspond to rare fluctuations in the far tails of this distribution where $I(s)$ becomes small. These bursts therefore represent non-Gaussian rare events that strongly influence the entropy dynamics of living systems.
From this perspective, the fluctuation theorem not only constrains the statistics of entropy production but also provides a quantitative framework for understanding the intermittent entropy bursts that shape the entropy bathtub dynamics across different biological life stages.

\subsection{Critical Dynamics of Aging and Death}\label{Aging}

Aging and death in living systems are often associated with critical slowing down and long-time memory effects in the underlying dynamics. These ingredients naturally suggest that aging is not merely a gradual deterioration but may instead represent a critical dynamical process in which the system approaches a tipping point. Beyond this point, the system can no longer sustain its non-equilibrium steady state. To describe this process we introduce a coarse-grained damage variable $D(t)$ representing the accumulated irreversible alterations in the organism, including oxidative damage to proteins, DNA mutations, and loss of cellular function. The dynamics of $D(t)$ result from the competition between damage accumulation and repair processes, both of which are influenced by the system's long-term memory.

Following the memory-modified Ginzburg--Landau dynamics for the collective order parameter $\Phi(t)$ introduced in Eq.~(\ref{GL_memory}),
we postulate that the damage variable obeys a generalized memory-dependent evolution equation \cite{Henkel2010}
\begin{equation}
\frac{dD(t)}{dt}=a_0(\mu-\mu_c)\int_0^t \gamma(t-s) D(s)\,ds-b\int_0^t \gamma(t-s) D(s)^3\,ds+\eta(t),
\label{damage}
\end{equation}
where $\eta(t)$ is a stochastic force whose correlations are governed by the colored-noise spectrum $S_\xi(\tau)$. The parameter $\mu$ acts as an age-related control parameter describing the gradual accumulation of stress in the organism. In general $\mu$ slowly evolves with time, 
$\mu(t)=\mu_0+rt$, reflecting the progressive biological aging of the system.

For the Keldysh framework, the linear response of the damage variable in frequency space is
characterized by the retarded Green function
\begin{equation} 
G^R(\omega)= \frac{1}{-i\omega+\tilde{\gamma}(\omega)-\Sigma^R(\omega)}.
\label{Retarded_GF}
\end{equation}
It is given from the Dyson equation of Eq.~\eqref{Dyson_equation} with $G_0^R(\omega)^{-1}=-i\omega+\tilde{\gamma}(\omega)$,
corresponding to the bare response of the generalized Langevin dynamics. Here $\tilde{\gamma}(\omega)$ shows the memory-dependent repair dynamics and $\Sigma^R(\omega)$ denotes the effective self-energy generated by environmental fluctuations and collective interactions. The relaxation time of the damage variable is determined by the pole of $G^R(\omega)$, giving
\begin{equation}
\tau_{\mathrm{rel}}^{-1}=\mathrm{Re}\!\left[\tilde{\gamma}(0)-\Sigma^R(0)\right].
\end{equation}
The critical point $\mu_c$ therefore corresponds to the condition where this effective relaxation rate vanishes, signaling the loss of dynamical stability of the living state.

In Eq.~\eqref{damage}, the coefficient of the linear term changes sign at the critical point $\mu_c$, marking the threshold at which damage accumulation overcomes repair mechanisms. In this sense $\mu_c$ represents a dynamical tipping point separating a stable living phase from an unstable regime of runaway damage. When environmental fluctuations exhibit long-range temporal correlations, the memory kernel decays algebraically
\begin{equation}
\gamma(\tau) \sim \tau^{-\theta},
\end{equation}
with $0<\theta<1$. Such long-tailed memory kernels generate non-Markovian relaxation dynamics. The exponent $\theta$ controls the persistence of memory in the repair dynamics. Smaller $\theta$ corresponds to a more slowly decaying kernel, stronger long-time memory, and therefore slower relaxation of the damage variable. By contrast, larger $\theta$ leads to a faster decay of memory and a dynamics closer to the Markovian limit. Accordingly, decreasing $\theta$ is expected to enhance low-frequency fluctuations, broaden the late-time tails in $D(t)$, and strengthen critical slowing down near $\mu_c$, whereas increasing $\theta$ weakens these effects and yields a faster approach to the stable or damaged steady state. As discussed in Eq.~\eqref{corr_time_diverge}, the system exhibits critical scaling near the transition point $\mu_c$. The relaxation time diverges as
\begin{equation}
\tau_{\mathrm{rel}}\sim|\mu-\mu_c|^{-\nu},
\label{critical_slowing} 
\end{equation}
indicating the onset of critical slowing down. For the aging dynamics considered here, the exponent is related to the memory parameter $\theta$ through $\nu \propto (1-\theta)^{-1}$. Biologically this implies that the organism becomes progressively less capable of recovering from perturbations as it approaches the end of life. 
For $\mu>\mu_c$ the system evolves toward a damaged steady state with finite damage amplitude
\begin{equation}
D_{\mathrm{st}}\sim(\mu-\mu_c)^{1/2},
\end{equation}
analogous to the order parameter in a second-order phase transition. This state corresponds to the elevated entropy plateau sometimes
observed in very old organisms prior to terminal decline.
The entropy production rate near the critical point exhibits singular behavior is closely related to the critical slowing down of relaxation dynamics, which follows the scaling of Eq.~\eqref{entropy_production}. This divergence reflects intense bursts of energy dissipation as the organism approaches dynamical instability.

\begin{figure}[tbp]
 \centering 
 \includegraphics[width=0.6\textwidth]{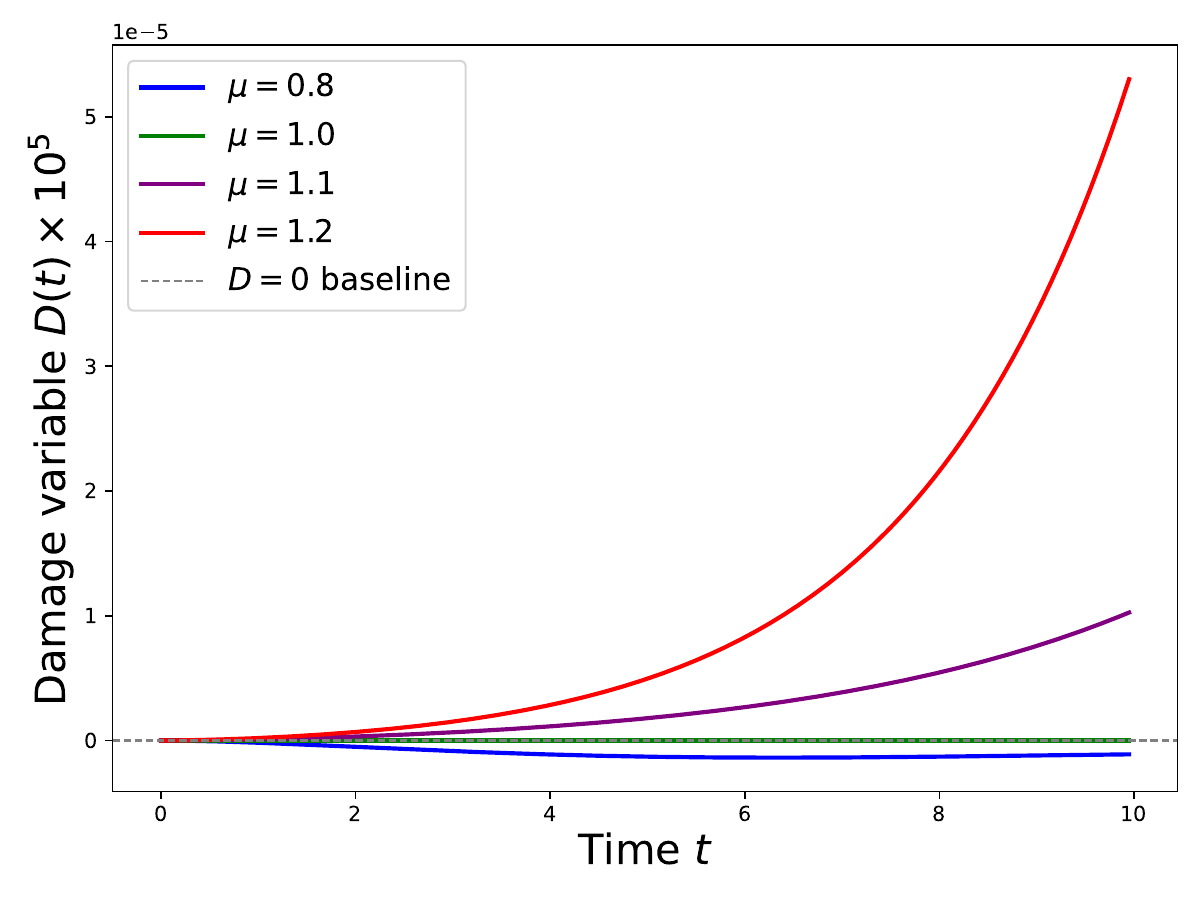}
 \caption{Time evolution of the damage variable $D(t)$ from Eq.~(\ref{damage}) with a power‑law memory kernel ($\theta = 0.5$). Below the critical point ($\mu = 0.8$, blue), damage decays to zero. At the critical point ($\mu = 1.0$, green), damage grows extremely slowly, exhibiting critical slowing down. Above the critical point ($\mu = 1.1$, purple; $\mu = 1.2$, red), damage accelerates and eventually saturates at a finite steady state. The gray dashed line marks $D=0$.} \label{fig:damage} 
\end{figure}

To illustrate these ideas we numerically integrate Eq.~(\ref{damage}) using a power-law memory kernel
\begin{equation}
\gamma(\tau) = (\tau+\varepsilon)^{-\theta},
\end{equation}
with $\theta=0.5$ and a small cutoff $\varepsilon=10^{-3}$ that regularizes the short-time behavior. The intermediate value of $\theta=0.5$ is adopted here to illustrate a regime with substantial long-time memory. Smaller values of $\theta$ would produce even slower relaxation and longer temporal tails, whereas larger values would move the dynamics closer to exponential, near-Markovian relaxation. The parameters are chosen as
$a_0=1$, $b=1$, and $\mu_c=1$. Figure~\ref{fig:damage} shows the resulting time evolution of $D(t)$ for four values of the control parameter. Below the critical point ($\mu<\mu_c$) damage decays and the system remains dynamically stable. At the critical point the relaxation becomes extremely slow, reflecting the divergence of the time scale in Eq.~\eqref{critical_slowing}. Above the critical point the linear instability drives rapid damage growth that is eventually stabilized by the nonlinear term, producing a finite damaged state.

These results connect directly with the non-Markovian entropy bathtub discussed in Section~\ref{Bathtub}. Long memory times amplify low-frequency fluctuations, leading to the accelerated entropy increase observed in late-life dynamics. Within this framework aging can therefore be interpreted as the gradual approach to a dynamical critical point controlled by accumulated damage and long-range memory. Death corresponds to the loss of dynamical stability beyond this point, after which damage grows irreversibly and the system rapidly relaxes toward thermodynamic equilibrium.

\subsection{Non-Markovian Response Functions to External Perturbations}\label{Response}

Understanding the response of living systems to external perturbations, such as environmental stress, therapeutic interventions, and sudden damage events that disturb the system from its steady state, is essential for characterizing their nonequilibrium dynamics. The response of a system with memory is described by a non-Markovian response function that generalizes the standard susceptibility of equilibrium statistical mechanics. For a weak external perturbation $f(t)$ coupled linearly to a dynamical variable $x(t)$, the response of the system is described by the linear response relation \cite{Kubo1966}
\begin{equation}
\langle \delta x(t) \rangle=\int_{-\infty}^{t}\chi(t-s)\,f(s)\,ds ,
\end{equation}
where $\chi(\tau)$ is the response function. Causality requires $\chi(\tau)=0$ for $\tau<0$. Physically, $\chi(\tau)$ quantifies how strongly a perturbation applied at time $t-\tau$ influences the observable at time $t$, and thus characterizes the temporal memory of the system. The Fourier transform of $\chi(\tau)$ yields the frequency-dependent susceptibility $\chi(\omega)$ discussed previously in Sec.~\ref{Noise}.

For systems governed by the generalized Langevin equation introduced in Eq.~\eqref{langevin}, the overdamped dynamics relevant for many biological processes can be written as \cite{Mori1965}
\begin{equation}
\dot{x}(t)=-\int_0^t \gamma(t-s)\,x(s)\,ds+f(t)+\xi(t),
\end{equation}
where $\gamma(\tau)$ is the memory kernel and $\xi(t)$ represents colored environmental noise.Linearizing the dynamics with respect to the perturbation yields the response function and its frequency-domain form, encoding the modification of the system's susceptibility by environmental memory. Taking the Laplace transform yields
\begin{equation}
\tilde{\chi}(s)=\frac{1}{s+\tilde{\gamma}(s)} ,
\end{equation}
where $\tilde{\gamma}(s)$ is the Laplace transform of the memory kernel. The susceptibility determines the linear response of the system and in this description the temporal structure of the response is directly controlled by the memory kernel $\gamma(\tau)$. If the kernel decays rapidly, the response reduces to the familiar exponential relaxation of Markovian systems. In contrast, long-tailed kernels produce slow power-law relaxation or oscillatory responses, implying that perturbations may influence the system for long periods of time. Within the Keldysh field-theoretical framework, the response function $\chi(\omega)$ corresponds to the retarded Green function $G^R(\omega)$ of the system, which has been given in Eq.~\eqref{Retarded_GF}.

For discrete-state systems described by the master equation of Eq.~\eqref{GME}, the linear response can be expressed in matrix form. Let $\mathbf{p}(t)$ denote the probability vector of system states. A small perturbation $f(t)$ that modifies the transition kernel leads to \cite{Zwanzig2001}
\begin{equation}
\delta \tilde{\mathbf{p}}(s)=\tilde{\mathbf{R}}(s)\,\tilde{f}(s),
\end{equation}
with the response matrix
\begin{equation}
\tilde{\mathbf{R}}(s)=\left(s\mathbf{I}-\tilde{\gamma}(s)\right)^{-1}\frac{\partial \tilde{\gamma}(s)}{\partial f}\mathbf{p}^{\mathrm{st}} ,
\end{equation}
where $\mathbf{p}^{\mathrm{st}}$ denotes the stationary distribution. The poles of $\tilde{\mathbf{R}}(s)$ move toward the imaginary axis, producing divergent relaxation times and enhanced sensitivity to perturbations, which determine the relaxation time scales of the system.

The response function is directly connected to the generalized fluctuation–dissipation ratio $X_{AB}(\omega)$, which equals unity in equilibrium and deviates from unity in nonequilibrium systems due to external driving and memory effects. Within the entropy bathtub framework, $\chi(\tau)$ describes the response of the system to external perturbations and its subsequent relaxation toward the steady state. Transient perturbations, such as acute disturbances, induce deviations that decay according to the response function, while the presence of memory leads to long temporal tails reflecting persistent environmental correlations. Sustained or irreversible perturbations may shift the system toward a new nonequilibrium steady state characterized by enhanced entropy production. Such behavior can already arise within the Markovian entropy-bathtub framework under a permanent change of driving force, while the non-Markovian extension considered here modifies the transient pathway, relaxation times, and stability properties associated with the approach to that new steady state. The non-Markovian response function therefore provides a quantitative link between the microscopic memory kernel and macroscopic observables, connecting spontaneous fluctuations, entropy production, and the dynamical responses of living systems to environmental or stochastic perturbations.

\subsection{Evolution of System-Environment Entanglement Entropy}\label{Evolution}

Living systems are intrinsically open systems that continuously exchange energy and information with their environment. These interactions generate correlations that store information between the system and its surroundings. To fully characterize this information exchange, it is therefore necessary to quantify the buildup and decay of system--environment correlations during the system's evolution. A natural measure of these correlations is the entanglement entropy and the associated mutual information between the system and its environment. Studying their time evolution provides an information-theoretic perspective on biological organization and reveals how living systems maintain order by continuously generating and sustaining correlations with the external world.

The state of the system is described by a reduced density matrix $\rho_S(t)$. 
Its quantum entropy is the von Neumann entropy \cite{Neumann1955}
\begin{equation}
S_{\mathrm{vN}}(t) = -\Tr[\rho_S(t)\ln\rho_S(t)] .
\end{equation}
When the density matrix is diagonal in the chosen basis, this expression reduces to the classical Shannon entropy  $S_{\mathrm{sys}}(t)$.
In general, $S_{\mathrm{vN}}(t)$ is smaller than $S_{\mathrm{sys}}(t)$ because the off-diagonal elements of $\rho_S$ encode coherent superpositions that carry additional information. 
The difference $S_{\mathrm{sys}}-S_{\mathrm{vN}}$ therefore quantifies the amount of quantum coherence present in the system.

Interactions between the system and its environment generate not only dissipation and stochastic noise but also correlations and entanglement. 
If the combined system and environment initially form a pure state, their joint evolution preserves purity while distributing quantum correlations between them. 
These correlations are quantified by the mutual information
\begin{equation}
I_{S:E}(t) = S_{\mathrm{vN}}(t) + S_{\mathrm{vN}}^{(E)}(t) - S_{\mathrm{vN}}^{(SE)}(t),
\end{equation}
which measures the total amount of information shared between the system and the environment. 
In the classical limit this reduces to the standard mutual information between random variables, while in the quantum regime it includes genuine entanglement.

In the weak-coupling limit, the mutual information can be approximated using the second-order cumulant expansion of the influence functional. 
This leads to
\begin{equation}
I_{S:E}(t)\approx\frac12\int_0^t ds\int_0^t du\;\chi(s-u)\, C(s-u).
\end{equation}
This expression shows that mutual information grows when the system's response aligns with environmental correlations, allowing the system to effectively ``learn'' about its surroundings and store information through correlated states.

By the Keldysh response function $\chi(t)$ corresponding to the retarded Green function $G^R(t)$, the environmental correlation function $C(t)$ is related to the Keldysh component $G^K(t)$ through
\begin{equation}
C(t)=\frac{i}{2}G^K(t).
\end{equation}
Using these relations, the mutual information can be written in terms of the Keldysh Green functions as
\begin{equation}
I_{S:E}(t)\approx \frac{i}{4}\int_0^t ds\int_0^t du\; G^R(s-u)G^K(s-u).
\end{equation}
This representation shows that the buildup of system–environment correlations is governed by the interplay between dynamical response and environmental fluctuations encoded in the Keldysh Green functions.

\begin{figure}[tbp]
\centering
\includegraphics[width=1\textwidth]{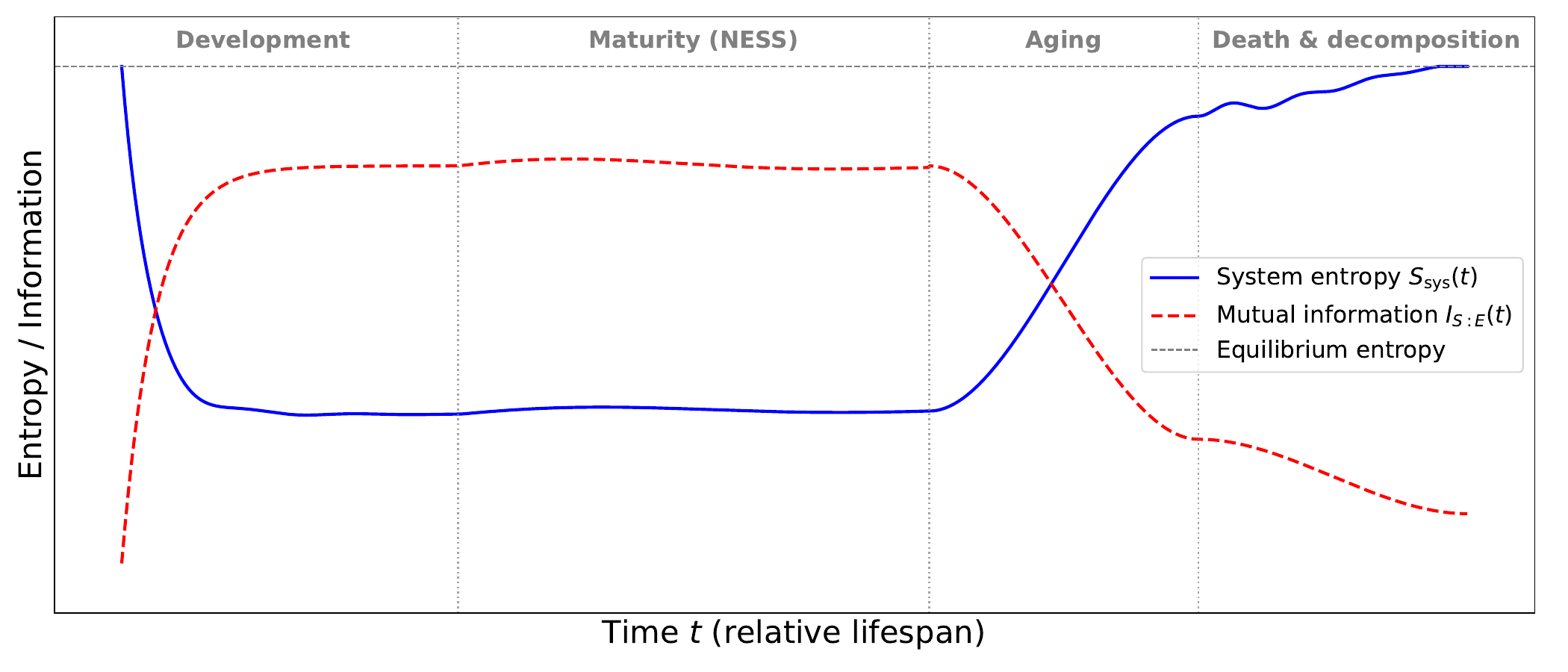}
\caption{Comparison of system entropy $S_{\mathrm{sys}}(t)$ (blue) and system–environment mutual information $I_{S:E}(t)$ (red) during the lifespan. The vertical dotted lines separate the developmental, maturity, aging, and death phases. During development, $S_{\mathrm{sys}}$ decreases while $I_{S:E}$ increases, reflecting the buildup of structured correlations with the environment. In the mature phase both quantities stabilize, while during aging the mutual information begins to decay before a significant rise in entropy. In the death phase $I_{S:E}$ rapidly drops while $S_{\mathrm{sys}}$ approaches the equilibrium entropy (horizontal dashed line).}
\label{Entanglement}
\end{figure}

Figure~\ref{Entanglement} shows the evolution of the system entropy $S_{\mathrm{sys}}(t)$ of entropy bathtub (blue solid curve) and the system–environment mutual information $I_{S:E}(t)$ of entanglement entropy (red dashed curve), with important differences. The curves are obtained by numerically solving the generalized master equation with the same exponential memory kernel and life-stage divisions used in the model of entropy bathtub in Section \ref{Bathtub}.

During the developmental phase, the system entropy decreases as biological order emerges, while the mutual information increases sharply. 
This indicates that the system forms increasingly strong correlations with its environment, effectively embedding structural information into environmental degrees of freedom. 
For example, a protein folding into its native conformation becomes correlated with surrounding solvent molecules and ions, creating a persistent information signature.
In the mature phase, the mutual information remains high and fluctuates around a plateau, reflecting continuous information exchange necessary for maintaining homeostasis. 
The system entropy remains low and stable, while the memory-induced fluctuations discussed earlier produce small oscillations around the steady state.
During aging, the mutual information begins to decay even before a noticeable increase in the system entropy occurs. 
This indicates that the weakening of system–environment correlations precedes the visible growth of internal disorder. 
Such a loss of correlations reflects the gradual degradation of functional organization and the reduced ability of the system to maintain coherent interactions with its surroundings.
Finally, in the death and decomposition phase, the mutual information drops rapidly toward a small residual value, indicating the near-complete loss of correlations between the system and the environment. At the same time, the system entropy approaches the equilibrium value as the system relaxes toward thermodynamic equilibrium.

From the Keldysh perspective, entropy production and mutual information represent two complementary aspects of the same nonequilibrium dynamics: the imbalance between fluctuations and dissipation generates both irreversible entropy and system–environment correlations. The evolution of system–environment entanglement entropy therefore provides the information-theoretic completion of the thermodynamic framework developed throughout this work. It unifies the microscopic memory kernel, the entropy bathtub, fluctuation–dissipation relations, response functions, and critical dynamics into a single coherent picture: living systems are open nonequilibrium systems that maintain order by continuously exchanging information with their environment. The trajectory of life, from development through maturity to aging and death, is encoded not only in the system's entropy but also in the correlations it shares with the surrounding world.

\section{Conclusion}\label{Conclusion}

In this work, we have developed a non-Markovian nonequilibrium extension of the entropy bathtub model within the Keldysh field-theoretical framework. By treating the system and its environment on equal footing, the formalism naturally incorporates environmental correlations, memory effects, and colored fluctuations arising from microscopic couplings. Within this framework, the entropy trajectory of a living system emerges from the interplay between dissipation, fluctuations, and system--environment interactions, yielding a thermodynamically consistent coarse-grained description of an illustrative life-history trajectory, from ordering and quasi-steady behavior to relaxation.

A central contribution of the present work is to provide a unified microscopic framework that connects system-environment coupling, memory kernels, generalized Langevin dynamics, non-Markovian master equations, entropy production, and fluctuation-dissipation violation within a single formalism. In contrast to previous Markovian formulations, the Keldysh framework allows the derivation of nonlocal memory kernels and generalized fluctuation-dissipation relations that govern the evolution of entropy in nonequilibrium biological systems. This enables the analysis of the influence of colored noise, non-Gaussian fluctuations, and many-body interactions on the classical entropy bathtub profile. Furthermore, the theory provides explicit expressions for non-Markovian entropy production and allows the investigation of critical dynamical features associated with slow relaxation and the breakdown of nonequilibrium steady states.

Beyond its conceptual implications, the framework developed here opens several promising directions for future research. In particular, the present formulation could be extended to incorporate realistic biochemical reaction networks, spatially extended systems, and active matter models that capture the collective behavior of living tissues. The connection between entropy evolution and experimentally measurable quantities, such as metabolic energy fluxes or fluctuation-response spectra, also remains an important direction for bridging theory and biological observations.

More broadly, the non-Markovian entropy bathtub framework suggests that entropy evolution in living systems may be interpreted as an emergent thermodynamic trajectory shaped by long-range temporal correlations and environmental interactions at a coarse-grained level. Future work combining nonequilibrium field theory, stochastic thermodynamics, and biological theory may therefore provide deeper insight into the universal thermodynamic principles underlying memory-driven organization, relaxation, and biological variability.

\section*{List of Symbols}

\begin{description}

\item[$t$] Time.

\item[$\tau$] Time difference, memory time, or observation time depending on context.

\item[$\omega$] Angular frequency in Fourier space.

\item[$s$] Laplace transform variable.

\item[$\hbar$] Reduced Planck constant.

\item[$k_B$] Boltzmann constant.

\item[$T$] Environmental temperature.

\item[$\beta$] Inverse temperature, $\beta=1/T$.

\item[$H_{\mathrm{tot}}$] Total Hamiltonian of system plus environment.

\item[$H_S$] System Hamiltonian.

\item[$H_B$] Bath Hamiltonian.

\item[$H_I$] System--environment interaction Hamiltonian.

\item[$\ket{x}$] Discrete system state.

\item[$E_x$] Energy of system state $\ket{x}$.

\item[$b_k,\; b_k^\dagger$] Annihilation and creation operators of the $k$th bath mode.

\item[$\omega_k$] Frequency of the $k$th bath mode.

\item[$Q_k$] Coordinate (amplitude) of the $k$th environmental fluctuation mode.

\item[$P_k$] Conjugate momentum of the $k$th environmental fluctuation mode.

\item[$B_x$] Bath coupling operator associated with system state $x$.

\item[$g_{x,k}$] Coupling constant between system state $x$ and bath mode $k$.

\item[$g_x$] Effective system--bath coupling amplitude.

\item[$\rho_B$] Equilibrium density matrix of the bath.

\item[$C_{xx'}(\tau)$] Bath correlation function associated with states $x$ and $x'$.

\item[$J_{xx'}(\omega)$] Bath spectral density.

\item[{$\mathcal{F}[\phi^+,\phi^-]$}] Influence functional after integrating out bath degrees of freedom.

\item[$\phi_x^\pm$] Forward/backward Keldysh contour fields.

\item[$\mathcal{K}_{xx'}(\tau)$] Microscopic non-Markovian memory kernel in the influence functional.

\item[$p_x(t)$] Probability of finding the system in state $x$ at time $t$.

\item[$K_{xx'}(\tau)$] Coarse-grained transition memory kernel in the generalized master equation.

\item[$k_{xx'}$] Baseline transition rate from state $x'$ to state $x$.

\item[$\delta(t)$] Time-dependent external driving bias.

\item[$\eta_{\mathrm{dis}}(t)$] Transient disease or perturbation pulse during the mature phase.

\item[$S_{\mathrm{sys}}(t)$] System entropy, $S_{\mathrm{sys}}(t)=-\sum_x p_x(t)\ln p_x(t)$.

\item[$q(t)$] Coarse-grained dynamical coordinate in the generalized Langevin equation.

\item[$x(t)$] Generic dynamical observable.

\item[$\xi(t)$] Stochastic force representing environmental fluctuations.

\item[$S_\xi(t-t')$] Noise correlation function.

\item[$S_\xi(\omega)$] Noise power spectrum in the frequency domain.

\item[$\gamma(\tau)$] Memory-dependent friction kernel.

\item[$\tilde{\gamma}(\omega)$] Fourier transform of the memory kernel.

\item[$\tilde{\gamma}(s)$] Laplace transform of the memory kernel.

\item[$\chi(\tau)$] Time-domain response function.

\item[$\chi(\omega)$] Frequency-domain susceptibility.

\item[$\tilde{\chi}(s)$] Laplace-domain response function.

\item[$V(q)$] Effective potential governing the system dynamics.

\item[$m$] Effective mass of the coarse-grained dynamical variable.

\item[$S_{AB}(t)$] Symmetrized correlation function of observables $A$ and $B$ in the time domain.

\item[$S_{AB}(\omega)$] Symmetrized correlation spectrum in the frequency domain.

\item[$\chi_{AB}^R(\omega)$] Retarded response function between observables $A$ and $B$.

\item[$\chi_{AB}''(\omega)$] Imaginary part of the retarded response function.

\item[$X_{AB}(\omega)$] Generalized fluctuation--dissipation ratio.

\item[$T_{\mathrm{ref}}$] Reference temperature used in the generalized fluctuation--dissipation relation.

\item[$T_{\mathrm{eff}}(\omega)$] Effective frequency-dependent temperature.

\item[$G^R(\omega)$] Retarded Green function.

\item[$G^A(\omega)$] Advanced Green function.

\item[$G^K(\omega)$] Keldysh Green function.

\item[$G_0^R(\omega)$] Bare retarded Green function.

\item[$\Sigma^R(\omega)$] Retarded self-energy.

\item[$\Sigma^K(\omega)$] Keldysh self-energy.

\item[$\Sigma''(\omega)$] Dissipative part of the retarded self-energy, $\Sigma''(\omega)=\mathrm{Im}\,\Sigma^R(\omega)$.

\item[$\Gamma$] Forward stochastic trajectory of the system.

\item[$\tilde{\Gamma}$] Time-reversed stochastic trajectory.

\item[{$\mathcal{P}[\xi]$}] Probability weight of a noise realization.

\item[{$\mathcal{P}[\Gamma]$}] Probability weight of a system trajectory.

\item[$\Delta S_{\mathrm{tot}}$] Total entropy production along a trajectory.

\item[$\langle \dot S_{\mathrm{tot}} \rangle$] Average entropy production rate.

\item[$I(s)$] Large-deviation rate function for entropy production fluctuations.

\item[$C_3(t_1,t_2,t_3)$] Third-order noise cumulant.

\item[$\Psi(t_1,t_2,t_3)$] Third-order response kernel.

\item[$N$] Number of interacting units in the many-body system.

\item[$P(\{x_i\},t)$] Joint probability distribution of the many-body system.

\item[$\mathcal{K}^{(i)}$] Single-site memory kernel in the many-body generalized master equation.

\item[$\mathcal{K}^{(ij)}$] Two-body interaction memory kernel in the many-body generalized master equation.

\item[$\phi(\mathbf r,t)$] Coarse-grained density field.

\item[$\Phi(t)$] Global order parameter.

\item[$U(\mathbf r-\mathbf r')$] Effective interaction potential.

\item[$D$] Diffusion coefficient in the coarse-grained field equation.

\item[$\zeta(t)$] Stochastic force in the order-parameter dynamics.

\item[$\mu$] Control parameter characterizing stress, coupling strength, density, or aging.

\item[$\mu_c$] Critical value of the control parameter.

\item[$\tau_c$] Characteristic memory or correlation time.

\item[$\tau_{\mathrm{rel}}$] Relaxation time.

\item[$\theta$] Exponent describing the long-time decay of the memory kernel.

\item[$\nu$] Critical exponent of the relaxation time.

\item[$\alpha$] Critical exponent of entropy production.

\item[$D(t)$] Damage variable describing accumulated biological deterioration.

\item[$D_{\mathrm{st}}$] Stationary damage amplitude above the critical point.

\item[$\rho_S(t)$] Reduced density matrix of the system.

\item[$S_{\mathrm{vN}}(t)$] von Neumann entropy of the system.

\item[$S_{\mathrm{vN}}^{(E)}(t)$] von Neumann entropy of the environment.

\item[$S_{\mathrm{vN}}^{(SE)}(t)$] von Neumann entropy of the combined system and environment.

\item[$I_{S:E}(t)$] Mutual information between system and environment.

\item[$C(t)$] Environmental correlation function entering the mutual-information expression.

\item[$\mathbf p(t)$] Probability vector of system states.

\item[$\mathbf p^{\mathrm{st}}$] Stationary-state probability vector.

\item[$\tilde{\mathbf R}(s)$] Response matrix in Laplace space.

\item[$\mathbf I$] Identity matrix.

\end{description}

\section*{Acknowledgement}
We thank Zhencheng Fu and Krzysztof Fornalski for helpful comments and insightful suggestions. This research was funded by Yunnan Provincial Department of Education Science Research Fund Project (Grant No. 2025J0942, 2026J0977), Yunnan Provincial Xiao Rui Expert Workstation (Grant No. 202605AF350031), 2025 Self-funded Science and Technology Projects of Chuxiong Prefecture (Grant No. cxzc2025004, cxzc2025008), Chuxiong Normal University Doctoral Research Initiation Fund Project (Grant No. BSQD\,\,\,   2407, BSQD2507),  and Dongying Science Development Fund (Grant No. DJB2023015).

\end{document}